\documentclass{article}

\usepackage[preprint]{neurips_2025}

\PassOptionsToPackage{numbers}{natbib}
\PassOptionsToPackage{table,dvipsnames}{xcolor}

\usepackage[utf8]{inputenc}
\usepackage[T1]{fontenc}

\usepackage{amsmath,amsfonts}
\usepackage{nicefrac,booktabs,microtype,url,graphicx}
\usepackage{xcolor}              

\usepackage{algorithm}
\usepackage[noend]{algpseudocode}

\usepackage{multirow,makecell,enumitem}

\usepackage{caption,subcaption,wrapfig}

\usepackage{tcolorbox,lipsum,soul,pifont,appendix,tocloft,
            circledsteps,fontawesome,marvosym,float}

\usepackage{natbib}
\usepackage[
  pagebackref=true,
  breaklinks=true,
  colorlinks=true,
  citecolor=blue,
  linkcolor=red,
  bookmarks=false
]{hyperref}

\usepackage{amssymb}
\usepackage{array}

\definecolor{lightgray}{gray}{0.75}  

\definecolor{blue}{rgb}{0.21,0.49,0.74}
\definecolor{red}{rgb}{0.8, 0.2, 0.2}
\definecolor{green}{rgb}{0, 0.5, 0}
\definecolor{yellow}{RGB}{218, 160, 109}

\title{Automated Vehicles Should be Connected with Natural Language}
\author{
     Xiangbo Gao, Keshu Wu, Hao Zhang, Kexin Tian, Yang Zhou, Zhengzhong Tu\thanks{Corresponding Author: Zhengzhong Tu (\texttt{tzz@tamu.edu})}\\[2pt]
  Texas A\&M University
}

\begin{document}

\maketitle




\begin{abstract}
Multi-agent collaborative driving promises improvements in traffic safety and efficiency through collective perception and decision-making. However, existing communication media—including raw sensor data, neural network features, and perception results—suffer limitations in bandwidth efficiency, information completeness, and agent interoperability. Moreover, traditional approaches have largely ignored decision-level fusion, neglecting critical dimensions of collaborative driving. In this paper, we argue that addressing these challenges requires a transition from purely perception-oriented data exchanges to explicit intent and reasoning communication using \textbf{natural language}. Natural language balances semantic density and communication bandwidth, adapts flexibly to real-time conditions, and bridges heterogeneous agent platforms. By enabling the direct communication of intentions, rationales, and decisions, it transforms collaborative driving from reactive perception-data sharing into proactive coordination, advancing safety, efficiency, and transparency in intelligent transportation systems.

\end{abstract}



\section{Introduction}

\label{sec:intro}

Recent advances in autonomous driving have demonstrated that multi-agent collaboration~\cite{liu2023towards} significantly enhances both safety and efficiency compared to single-vehicle operations, primarily through real-time information sharing and intention communication~\cite{ali2021efficient, zhang2024multi, gao2025airv2x, wang2025generative}. 
This collaborative approach has become increasingly crucial as autonomous vehicles navigate complex environments where interaction with other traffic participants is inevitable and constant~\cite{wu2025ai2,wu2024hypergraph, tian2025physically, pu5019798optimal, li4940014nonlinear, godbole2025drama}. 
However, the selection of an appropriate communication medium—one that balances information richness, bandwidth efficiency, and cross-platform compatibility—remains a critical challenge in the field.

A key element of multi-agent collaboration is the medium used for inter-vehicle communication. 
Researchers have proposed various modalities for exchanging information, including raw sensor data~\cite{chen2019cooper, marvasti2020cooperative}, neural network features~\cite{liu2020when2com, wang2020v2vnet}, and downstream task results~\cite{melotti2020multimodal, fu2020depth, zeng2020dsdnet, shi2022vips, glaser2023we}. Despite their utility, each of these communication media suffers one or more critical drawbacks from high communication bandwidth requirements, fails to accommodate the inherent heterogeneities across agents, the loss the critical contextual information, and lacks the support of decision-level collaboration. These limitations become particularly apparent in scenarios requiring rapid negotiation and decision-making among multiple agents, such as unsignalized intersections, highway merging, and unexpected road hazards~\cite{kosuru2023advancements, li2024sequencing, li2025simulating}. In such cases, for safe and efficient navigation, communication is not just the perception but also the reasoning processes and the intended actions.

Transportation systems exist ultimately to serve people, drivers, passengers, and pedestrians, so their most natural “language” should be human language itself. To address current limitations in V2X, we propose human natural language as a universal communication media for multi-agent collaborative driving. Unlike raw sensor or learned feature exchanges, language is immediately interpretable by humans and by any machine equipped with a shared ontology, ensuring seamless interoperability across heterogeneous vehicles and infrastructure. It also endows machines with human-like reasoning and negotiation abilities~\cite{liu2025colmdriver, cui2025talking, gao2025langcoop}—vehicles can explain intentions (\texttt{``I’m yielding because I detected a stalled car”}) and coordinate complex maneuvers proactively. Finally, the recent surge in large vision language models (LVLMs) enables driving agents to ground linguistic messages in visual context, delivering expert-level decision-making and a more holistic understanding of the environment~\cite{xing2025openemma, sima2024drivelm, gopalkrishnan2024multi, nie2024reason2drive, tian2025nuscenes, zhou2024vision}.

\begin{tcolorbox}[colback=gray!5!white, colframe=gray!75!black, 
title=Speak Human: V2X Communication]
If transportation is human-centric, then V2X communication must be too.
\end{tcolorbox}

\section{Pros and Cons of Existing Communication Media}

\paragraph{Raw Sensor Data.} 
Raw sensor data communication~\cite{gao2018object, chen2019cooper, arnold2020cooperative} is also named early collaboration, which shares complete environmental information captured by agents' sensors including raw LiDAR \& Radar point clouds, surround-view images, depth maps, high definition maps, and others~\cite{zhang2025anticipatory,zhang2022temporal}. This approach maximally preserves raw data, enabling agents to perform custom processing tailored to their specific needs. However, its implementation faces significant challenges due to \textbf{extreme bandwidth requirements}~\cite{hu2024collaborative}. In reality, most of the sensor information can barely help other connected agents. For example, transmitting 6 uncompressed 4K surround-view images with no critical actors being captured wastes both the communication bandwidth and computation time for all connected agents.

\paragraph{Perception Results.}
Perception result communication, or late collaboration, involves sharing processed outputs such as object detections~\cite{melotti2020multimodal, fu2020depth, zeng2020dsdnet, shi2022vips, glaser2023we, xu2023model, su2023uncertainty, su2024collaborative} (bounding boxes with class labels, positions, dimensions, and orientations), occupancy grids~\cite{tian2023occ3d, kalble2024accurate} (discretized representations of free and occupied space), or semantic segmentations~\cite{peng2023bevsegformer, gu2024clft, li2024sscbench}. Bounding boxes provide compact object-level information essential for collision avoidance and trajectory planning, while occupancy grids offer spatial understanding of navigable areas regardless of object classification. These representations deliver interpretable information about environmental elements with clear semantic meaning, enabling straightforward integration into recipient vehicles' planning systems without requiring extensive computational resources. The primary limitations include potential \textbf{task misalignment}. For example, occupancy grid predictions cannot be directly fused with bounding box results. Besides, \textbf{information loss} through abstraction is inevitable in late collaboration. For instance, a detection miss or false positive can cause severe results with bare possibility to correct.

\paragraph{Neural Network Features.}
Neural feature sharing, also known as intermediate collaboration, represents a middle-ground approach where vehicles exchange intermediate representations extracted from their perception models~\cite{hu2022where2comm, wang2020v2vnet, xu2022v2x, wangcocmt, xu2023cobevt}. The features contain compressed environmental understanding that can be integrated into recipient vehicles' perception systems. This method significantly reduces data volume while preserving information richness; therefore, it was once considered the most promising collaboration fusion medium and was extensively researched. However, these methods face the significant challenge of \textbf{heterogeneity issues}—heterogeneous agents are non-collaborative. To address this issue, HEAL~\cite{lu2024extensible} and STAMP~\cite{gao2025stamp} use backward alignment and collaborative feature alignment methods, respectively, to alleviate the heterogeneity among agents. Despite that these methods experimentally show that they solve the heterogeneity issue, the solution requires a significant amount of extra training and maintenance, making the overall system highly complex and unstable.

In addition to the aforementioned difficulties, researchers are also facing challenges including decision-level communication, scenario variability, and transparency and trustworthiness concerns, which will be further discussed in the following section.

\section{Core Challenges of Multi-agent Collaborative Driving}
\label{sec:challenges}

Multi-agent collaborative driving represents a paradigm shift in intelligent transportation systems, wherein various agents—vehicles~\cite{xu2022opv2v,xu2023v2v4real}, roadside units (RSUs)~\cite{xiang2024v2x}, unmanned aerial vehicles (UAVs)~\cite{wang2024uvcpnet}, mobile robots, and even pedestrians equipped with smart devices—work in concert to enhance overall traffic safety and efficiency. However, this promising framework faces several challenges that must be addressed to realize its full potential. This section examines the core challenges inhibiting effective multi-agent collaboration in intelligent driving systems.

\begin{figure}[ht]
    \centering
    \includegraphics[width=\linewidth]{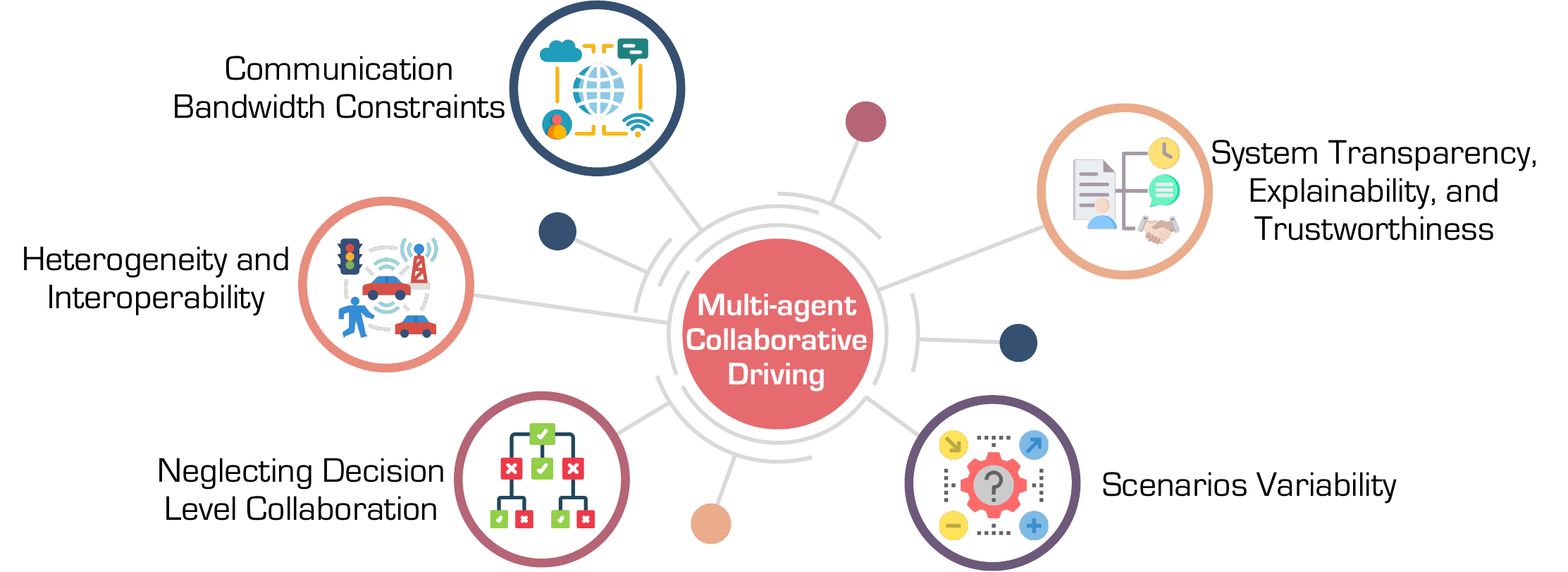}
    \caption{Core challenges of multi-agent collaborative driving.}
    \label{fig:multi_agent}
\end{figure}

\subsection{Communication Bandwidth Constraints}


\begin{table}[ht]
\centering
\setlength{\tabcolsep}{9.2pt}
\caption{Bandwidth Limitation and Latency Comparison of Widely Used V2X Communication Devices}
\begin{tabular}{lllll}
\toprule
\textbf{Communication Device} & \textbf{Bandwidth} & \textbf{Latency} \\
\midrule
DSRC  & 3–27 Mbps & 1–2 ms (light traffic)\\
LTE-V2X  & 50-100 Mbps & 10–100 ms \\
5G-V2X  & 500-3000 Mbps & 3–10 ms \\
\bottomrule
\end{tabular}
\label{tab:v2x_comparison}
\end{table}

The foundation of collaborative driving systems rests on reliable, high-throughput communication technologies to facilitate real-time information exchange. Current Vehicle-to-Everything (V2X) technologies exhibit varying capabilities in terms of bandwidth capacity and communication architectures.
DSRC (IEEE 802.11p) enables direct V2V and V2I communication over the 5.9\,GHz ITS band with 10\,MHz channels, supporting data rates between 3 and 27\,Mbps~\cite{kenney2011dsrc}. LTE-V2X (Release 14) introduces an enhanced PC5 sidelink interface with improved spectral efficiency, reaching up to 100\,Mbps in 20\,MHz channels under optimal conditions. The newer 5G-V2X (NR-V2X, Release 16+) substantially increases performance capabilities, supporting up to 100\,MHz bandwidth in sub-6\,GHz bands and multi-gigabit rates in mmWave bands, alongside ultra-reliable low-latency communication (URLLC) with latencies as low as 3–10\,ms~\cite{5gaa2020whitepaper}. Table~\ref{tab:v2x_comparison} summarizes these communication specifications.


\begin{figure}[ht]
  \centering
  \includegraphics[width=0.65\textwidth]{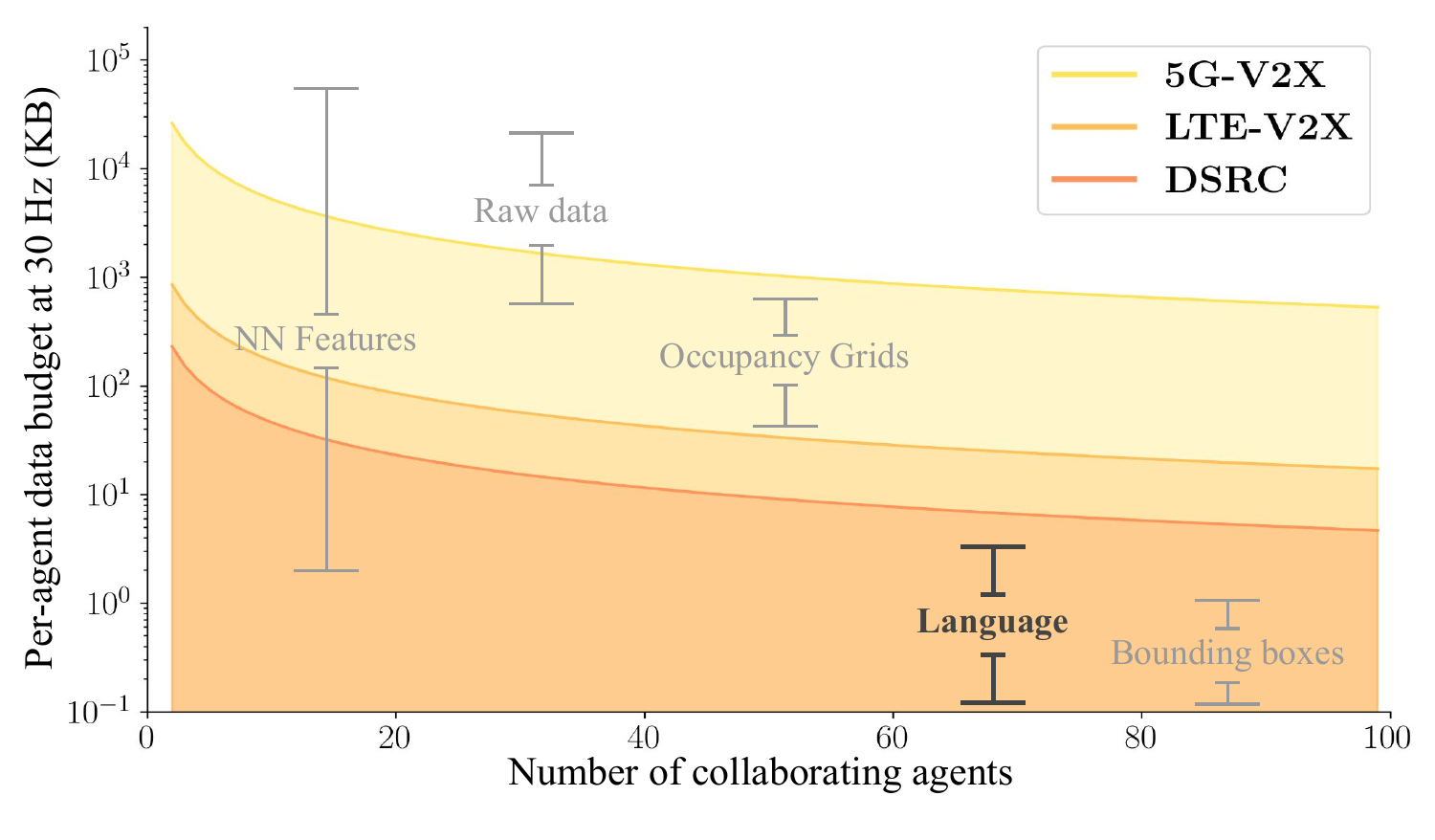}
  \caption{Per-agent data budget decreases significantly as the number of connected agents increases in collaborative driving systems. Data budgets are shown in kilobytes. Language and bounding box messages remain efficient across all three mainstream communication protocols, whereas neural network features, raw sensor data, and occupancy grids greatly exceed practical bandwidth limits.}
  \label{fig:comm_lim}
\end{figure}

Despite these technological advancements, bandwidth limitations remain a critical bottleneck for real-time collaborative driving, particularly in dense urban scenarios where the number of connected agents can grow substantially. As illustrated in Figure~\ref{fig:comm_lim}, the per-agent data budget diminishes dramatically as the number of participating agents increases under any given communication protocol. This severe constraint renders the transmission of raw sensor data (such as uncompressed LiDAR point clouds or camera images), occupancy grids, or extensive neural network features impractical for large-scale deployments. While some studies have experimented with transmitting only objects' bounding boxes to conserve bandwidth, this approach often results in suboptimal downstream task performance due to the loss of critical environmental information. Consequently, research is increasingly exploring more efficient information encoding methods, including natural language descriptions, which offer high semantic density with relatively low bandwidth consumption.

\subsection{Heterogeneity and Interoperability Challenges in A Complete V2X System}

In real-world collaborative driving scenarios, the participating agents exhibit heterogeneity across multiple dimensions, creating interoperability challenges~\cite{gao2025stamp, wang2025generative}. This heterogeneity manifests in several critical forms that greatly affect the system's collaborative capabilities.

One obvious form of heterogeneity appears in the \textbf{hardware and software configurations} of different vehicles. Vehicles from different manufacturers are often equipped with diverse sensor suites and driving algorithms. This diversity leads to incompatible sensor inputs, neural network architectures, and feature representations, potentially resulting in system errors.
\textbf{Task heterogeneity} represents another challenge, wherein different vehicles might be optimized for different downstream objectives. For example, one agent might generate object bounding boxes while another outputs occupancy grid predictions. Despite serving the common goal of trajectory planning and control, these output formats are not inherently compatible and cannot be naively fused without additional processing layers.
While some forms of heterogeneity—such as completely incompatible data formats—result in obvious system failures and program crashes that are readily identifiable, other forms produce more subtle compatibility issues that degrade performance without causing overt failures. For instance, when two vehicles employ models with identical architectures but are \textbf{trained on slightly different datasets}~\cite{xu2023bridging}, their feature representations may exhibit subtle divergences that lead to unexpected behaviors when their outputs are combined. These "silent failures" are particularly problematic because they do not trigger explicit error conditions but can nonetheless compromise the integrity of collaborative perceptions and decisions.




\begin{tcolorbox}[colback=gray!5!white, colframe=gray!75!black, 
title=Heterogeneity Metaphor]

In a world where each tribe speaks a different tongue, even the most urgent truths are lost in translation. Though they may all point to the same storm on the horizon, their voices rise in dissonance, and none is understood by the others. Without a shared language, wisdom becomes noise, and collaboration dissolves into isolation.

\end{tcolorbox}

In addition to inter-vehicle collaborative perception, a complete V2X system contains messages of multiple forms. SAE J2735~\cite{SAEJ2735_2024} defines a rich V2X message set enabling real-time safety, coordination, and information sharing across vehicles, infrastructure, and personal devices. Table~\ref{tab:v2x_messages} lists some typical message categories defined by SAE J2735. According to SAE J2735, data of different categories have distinct data representations. For example, map and geometry are represented in graph structure; signal phase and timing messages encode current signal states and countdown timers as structured arrays of integer values; environmental and testing data are in natural language form; safety and cooperative signals are mostly binary flags; while probe and sensor data are represented in structured lists of floating values~\cite{wu2025digital, li2024howdoes}. Sharing and taking usage of this giant set containing messages of different representations is way beyond the existing research of heterogeneous sensor modalities or feature representations. Researchers are looking for a unified representation that is able to encode most of these diversified messages~\cite{wu2025v2x}.

\begin{table}[h]
\centering
\small
\caption{SAE J2735~\cite{SAEJ2735_2024} V2X Message Set.}
\label{tab:v2x_messages}
\begin{tabular}{p{4.3cm}p{8.7cm}}
\toprule
\textbf{Category} & \textbf{Example Messages} \\
\midrule
Safety Beacons & Basic Safety Message (BSM), Personal Safety Message (PSM) \\
Cooperative Signaling & Signal Request Message (SRM), Signal Status Message (SSM) \\
Map and Geometry & MapData, Road Geometry \& Attributes \\
Signal Phase and Timing & Signal Phase and Timing (SPaT) Message \\
Probe and Sensor Data & Probe Vehicle Data (PVD), Probe Data Management (PDM) \\
Environmental and Correction & Road Weather Message (RWM), RTCM Corrections \\
Traveler Alerts & Roadside Alert (RSA), Traveler Information Message (TIM) \\
Testing and Charging & TestMessage, Road User Charging Config/Report (RUCCM/RUCRM) \\
\bottomrule
\end{tabular}
\end{table}

\subsection{Neglecting Decision Level Collaboration}

Current collaborative driving frameworks predominantly emphasize perception-level fusion, exchanging sensor data, neural network features, or detection results. This singular focus overlooks critical decision-level collaboration, where vehicles explicitly communicate intended actions and decision rationales. Such oversight can lead to scenarios where vehicles are fully aware of each other's presence but lack understanding of mutual intentions, resulting in inefficient maneuvers or even dangerous interactions during complex tasks like intersection crossings or lane merging.

\begin{tcolorbox}[colback=gray!5!white, colframe=gray!75!black,
title={Perception is a Tool, Driving is the Goal}]
Perception is merely the means to an end; the ultimate goal of collaborative driving is safe, efficient, and coordinated vehicle behavior.
\end{tcolorbox}

Integrating decision-level communication, such as explicit trajectory plans or anticipated maneuvers, would significantly enhance cooperative driving by enabling proactive conflict resolution and context-aware decision-making~\cite{liu2025colmdriver,manzinger2017negotiation, gao2025langcoop}. Effective decision-level collaboration would allow vehicles to anticipate each other's movements, negotiate driable areas, coordinate their actions, and ensure smoother, safer navigation, especially in high-stakes, dynamic traffic environments.

\subsection{Scenarios Variability}

Effective multi-agent collaboration requires the transmission of information that is "enough" for performing downstream tasks. Insufficient information exchange leads to various performance degradations~\cite{liu2023towards, huang2023v2x}, while redundant data transmission not only consumes precious bandwidth resources but can also complicate the extraction of crucial information by introducing noise into the collaborative perception and decision-making process. However, the definition of what constitutes "enough" information varies depending on the scenario. Environmental conditions significantly influence sensor effectiveness; radar sensors, for instance, demonstrate superior performance in adverse weather conditions such as heavy rain, fog, or snow, whereas LiDAR provides broader visibility in low-light scenarios compared to RGB cameras~\cite{zhang2025virtual}. Additionally, environments like urban canyons, characterized by dense infrastructure and complex dynamics, inherently demand richer situational awareness than more predictable, open highway settings. Bandwidth availability itself fluctuates significantly due to factors like network congestion, varying communication ranges, and interference from physical obstructions or other wireless devices. As a result, any static strategy for prioritizing information is likely to become inadequate as environmental and communication conditions evolve. Addressing this challenge requires developing adaptive information-sharing mechanisms capable of dynamically responding to environmental changes and bandwidth constraints, thereby ensuring consistent robustness and performance across diverse operational scenarios.

\subsection{System Transparency, Explainability, and Trustworthiness}

The increasing complexity of multi-agent collaborative driving systems introduces significant challenges related to transparency~\cite{9836249}, explainability~\cite{malik2021collaborative}, and trustworthiness~\cite{xing2024autotrust}—all of which are essential for widespread adoption and regulatory approval. The state-of-the-art intermediate fusion methods, despite their superior downstream task accuracy, remain largely opaque to both the users and their developers, making the collaboration process not transparent. A practical example is that a Waymo driverless car stopped dead inside a construction zone, causing disruptions and creating hazards~\cite{waymo2023}. Such incidents reveal a limitation of conventional sensor-based communication: it fails to transparently communicate the vehicle's internal decision-making and reasoning processes to nearby human drivers or traffic controllers.

Researchers and engineers have to bear in mind that the explainability challenge extends beyond technical debugging to encompass human factors considerations. Autonomous driving technology ultimately serves human users, necessitating a human-centric approach to system design~\cite{xing2021toward}. Drivers, passengers, and other road users require intuitive understanding of vehicle behaviors and decision rationales to establish appropriate trust and effectively collaborate with autonomous systems.

\section{Natural Language as the Ideal Communication Medium}
Having examined the limitations of current communication approaches, we now present the case for why natural language represents an ideal medium for multi-agent collaborative driving. Natural language offers unique advantages in expressiveness, efficiency, interoperability, and human alignment that make it particularly well-suited for autonomous vehicle communication.

\begin{figure}[ht]
    \centering
    \includegraphics[width=\linewidth]{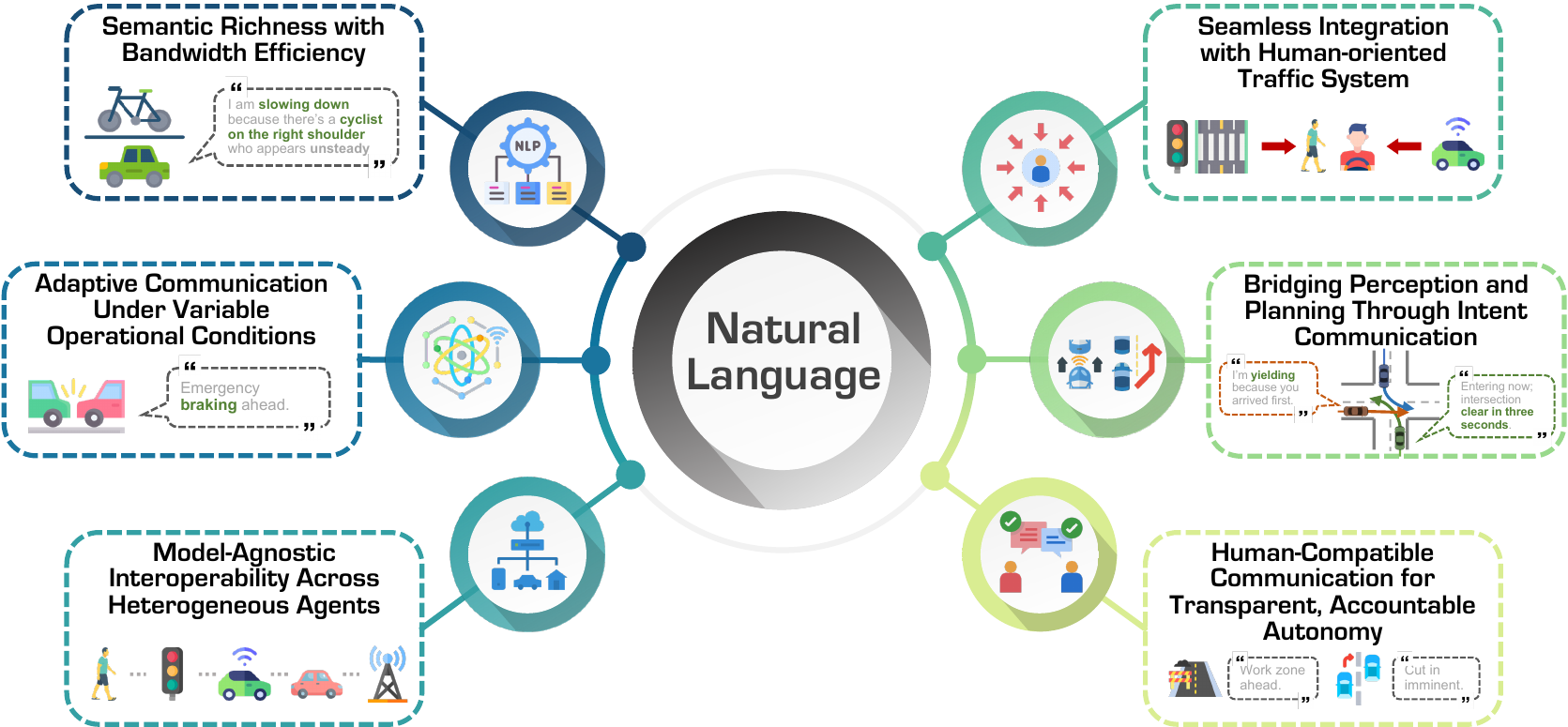}
    \caption{Natural language as the ideal communication medium. 
    }
\end{figure}

\subsection{Semantic Richness with Bandwidth Efficiency}
Natural language achieves a balance between information density and bandwidth efficiency. A concise textual description can convey complex environmental states, intentional stances, and reasoning processes in a few kilobytes of data~\cite{gao2025langcoop, luo2025senserag, you2024v2x, chiu2025v2v}, dramatically reducing bandwidth requirements compared to raw sensor sharing approaches. For instance, a message like "\texttt{I am slowing down because there's a cyclist on the right shoulder who appears unsteady}" communicates perception information (the presence and location of a cyclist), state assessment (the cyclist appears unsteady), intended action (slowing down), and causal reasoning (the relationship between the cyclist's state and the vehicle's decision) in less than 100 bytes. Conveying equivalent information through raw sensor data would require megabytes of LiDAR points, camera images, and trajectory predictions.

\subsection{Adaptive Communication Under Variable Conditions}
Natural language excels at dynamically scaling communication content based on real-time constraints and situational priorities. In bandwidth-limited environments (tunnels, rural areas, network congestion), vehicles can automatically compress messages to essential information: "\texttt{Emergency braking ahead}" rather than detailed scene descriptions~\cite{li2022federated}. Conversely, when bandwidth permits, the same linguistic framework allows for rich contextual details that enhance cooperative planning. This bandwidth adaptability occurs without protocol renegotiation, unlike fixed-format approaches that struggle with dynamic compression~\cite{hussein2022comprehensive}. Similarly, natural language communication seamlessly adapts to situation criticality—in normal driving, vehicles may exchange detailed intent and perception data, while emergency scenarios trigger prioritized, high-salience messages ("\texttt{Collision imminent, swerving right}") that command immediate attention across all communication channels~\cite{zhang2024context}. This inherent ability to scale communication complexity based on operational conditions makes language-based systems particularly robust across the diverse environments and scenarios autonomous vehicles must navigate.

\subsection{Model-Agnostic Interoperability Across Heterogeneous Agents}
Natural language provides a universal interface that enables interoperability among diverse autonomous systems without enforcing a common hardware stack or software architecture~\cite{driess2023palm}. Any vehicle equipped with a Large Vision-Language Model (LVLM) can generate and interpret language-based messages regardless of its sensor suite, perception pipeline, or planning algorithm~\cite{ahn2022can}. Crucially, "interoperable V2X" today means more than just vehicle-to-vehicle exchange; it also covers vehicle-to-infrastructure, vehicle-to-pedestrian, vehicle-to-drone, and future extensions such as vehicle-to-grounded-robot~\cite{alalewi20215g}. Each party needs a different slice of information—roadside units care about queue length, pedestrians care about crossing time, automated trucks care about bridge height—yet all can share the same linguistic channel~\cite{wei2022queue,abdi2024advancing}. Because natural language carries semantics in the words themselves, every agent can quickly parse a message, keep the fields that matter, and ignore the rest without custom protocol negotiation~\cite{xie2021deep}.

\subsection{Seamless Integration with Existing Human-oriented Traffic System}
There is an undeniable reality that the existing traffic system has been developed for nearly a century serving human users, from text-based street signage to advanced V2X linguistic instructions~\cite{traffic2009manual}. These systems were designed to accommodate the cognitive and linguistic capabilities of natural language speakers~\cite{ng2007cognitive}. Developing a parallel traffic system exclusively for autonomous agents would be neither effective nor economical~\cite{gopalakrishna2021impacts}. Moreover, purely isolated ecosystems are not suitable for large-scale human habitation~\cite{ullah2025assessing}. Autonomous traffic systems must therefore integrate with existing infrastructure without disrupting current traffic flows, which poses significant challenges.

By leveraging both vision and language as primary information modalities and communication media while utilizing LVLMs as the main intelligent "brain," emerging autonomous systems can seamlessly integrate with the existing traffic infrastructure~\cite{driess2023palm}. This approach allows autonomous vehicles to interpret the same signs, signals, and conventions that human drivers rely on, while also enabling them to communicate with each other and with infrastructure using the same linguistic framework that underpins our current traffic system~\cite{yao2024vision, gan2025planning}.

\subsection{Bridging Perception and Planning Through Explicit Intent Communication}
Natural language uniquely enables explicit communication of intentions, preferences, decision rationales, and negotiation~\cite{cui2025talking, gao2025langcoop}, bridging the gap between perception-level sensing~\cite{hu2024pragmatic} and planning-level action~\cite{cui2025talking}. With a shared linguistic channel, vehicles can exchange not only what they perceive but also what they will do next, why they will do it, and how they expect others to respond—fuel for context-aware trajectory planning, intent-aware prediction, and other multi-modal reasoning techniques. This capability shines in negotiation scenes where agents must resolve latent "games" in traffic flow~\cite{hua2024game,talebpour2015modeling}. At an uncontrolled intersection~\cite{fang2024cooperative,qin2024game}, for instance, cars can broadcast: "\texttt{I'm yielding because you arrived first,}" or "\texttt{Entering now; intersection clear in three seconds.}" Such explicit dialogue turns reactive inference into proactive consensus, boosting both safety and throughput~\cite{cui2025talking}.

Language also carries nuanced preferences and constraints that purely numeric protocols miss. A vehicle can declare urgency ("\texttt{Need the next right turn—medical emergency}~\cite{wu2020emergency,wu2021cooperative}") or physical limits ("\texttt{Cannot brake hard, fragile cargo onboard}"). Other agents then integrate this semantic context into context-aware prediction modules~\cite{zheng2024large,mo2024cross} and game-theoretic trajectory planners~\cite{zhao2024stackelberg,naidja2024gtp}, adjusting their own maneuvers to satisfy competing motives while preserving collective efficiency.

\subsection{Human-Compatible Communication for Transparent, Accountable Autonomy}
Perhaps the most distinctive advantage of natural language is its human compatibility—enabling transparent communication not only between autonomous vehicles but also with human drivers, pedestrians, and transportation authorities~\cite{cui2024board, xu2024drivegpt4}. Though exchanging detailed quantitative future motion data between human-driven vehicles and connected automated vehicles is unrealistic—human drivers cannot interpret precise trajectory predictions nor broadcast their own planned maneuvers—everyone benefits from sharing just enough context to anticipate hazards, such as "work zone ahead," "cut-in imminent," or "slowing for a stopped bus." Broadcasting these concise human language V2X messages allows connected vehicles to infer human driver intentions and upcoming events without accessing proprietary vehicle internals, while human drivers receive the same alerts via dashboards or smartphone apps. Natural language's semantic richness keeps each message lightweight (only a few bytes) yet immediately interpretable by all participants, including human-driven vehicles, automated vehicles, and remote fleet operators, bridging information asymmetries and enabling proactive, cooperative behavior in mixed traffic by providing the detailed context.

\section{Alternative Views and Potential Limitations}
While we advocate for natural language as a primary communication medium for collaborative driving, it is important to acknowledge alternative viewpoints and potential limitations of this approach. A balanced assessment strengthens our position by addressing legitimate concerns and identifying areas where complementary approaches may be valuable.

\subsection{Precision and Ambiguity Concerns}
\textbf{Alternative View:} Critics argue that natural language is inherently ambiguous and imprecise compared to structured numeric representations, potentially introducing safety risks in critical driving scenarios. For example, spatial descriptions like "nearby" or "approaching rapidly" lack the exact metric precision of coordinates and velocity vectors.

\textbf{Our Response:} While natural language can indeed be ambiguous in general contexts, \textbf{domain-specific language use in driving can achieve high precision} through consistent terminology and contextual grounding~\cite{xu2024drivegpt4, sima2024drivelm}. Messages can include specific metrics when needed (e.g., "braking to stop 10 meters before intersection") while maintaining the flexibility to express concepts difficult to capture in pure numeric form. Recent research shows that with appropriate training and prompting, LVLMs can achieve remarkably consistent interpretations of driving-related language~\cite{xing2025openemma, gao2025langcoop, luo2025v2x}, particularly when combined with spatial grounding. 

On the other hand, \textbf{we would like to question that if receiving information with high numerical precision is mandatory in driving}. Current V2X system is built based on the preliminary that each agent is able to perform basic actions, shared information is intended to provide cross validation or additional information that further improve the safety and efficiency. Taking ``pedestrian dart-out", a pedestrian runs out from a vision blind spot, one of the most critical application of V2X communincation as an example, An natural language text like: "\texttt{There is an pedestrian in front of the black SUV in your front right. Please be careful.}" is enough for reminding autonomous vehicles to slow down to avoid the collision. Another example is that a widely used V2X notification: "Frequent accidents ahead. Please drive with caution." is directly understandable and helpful for language-based autonomous vehicles. In comparison, converting such accidental alert into structural numerical data is either trivial nor necessary.

\subsection{Computational Efficiency and Latency}
\textbf{Alternative View:} Processing natural language requires computational resources that could be better allocated to core perception and planning tasks. The generation and interpretation of messages through LVLMs might introduce unacceptable latencies in time-critical scenarios.

\textbf{Our Response:} While LVLMs are indeed computationally intensive in their full form, \textbf{specialized models optimized for driving} can run with less resource requirements~\cite{dong2024generalizing, chiu2025v2v}. Besides, as the development of LVLMs, model compression algorithms, computing devices, it is promising that the driving-specialized LVLMs will become efficient enough for real-time driving purpose. Additionally, their is a tendency of using LVLMs in autonomous driving.\textbf{ It is the usage of LVLMs that causes the latency, not the communincation media itself}. If drive-specicalized LVLMs is encouraged, using natural language as communication media, as one of the most direction communication way for AVs that have already equipped with LVLMs, is also worth researching and should be encouraged. 

\subsection{Security and Trust Concerns}
\textbf{Alternative View:} As the natural language to be an univerisal communincation media, implicating any actors ``speaks" natural languages can easily manipulate the system? Would it more vulnerable to spoofing, semantic manipulation, or adversarial attacks than structured data formats with clear validation rules or fire walls.

\textbf{Our Response:} The improvement of the interoperability of the system inevitably creates or enlarge some security risks upon the tranditionally isolated system. Future researches should focus on finding new methods or optimizing the existing one to enhance the system security. Consider that natural language is a product of the mind, one possible approach is to imitating the human solutions towards security and trust concerns. For example, in the vehicle-to-pedestrian communication, vehicle should only strictly follow the instruction of people wearing police uniform with valid license, instead of a five-year-old kid.

\subsection{The Case for Hybrid Approaches}
\textbf{Alternative View:} Some researchers propose that optimal collaborative driving will require hybrid communication approaches that combine multiple types of data exchange with natural language.

\textbf{Our Response:} 

We acknowledge merit in this perspective. In particular scenarios, direct exchange of precise numeric data (e.g., GPS coordinates for path planning) may complement language-based communication. Our position is not that natural language should be the exclusive communication medium but rather the primary, universal protocol that provides an interoperable foundation across heterogeneous systems. Language can serve as the coordinating and contextualizing layer that gives meaning to any accompanying structured data, similar to how humans might share a map location while explaining its significance. This hybrid approach preserves the semantic richness and interoperability of language while incorporating the precision of structured data where beneficial.

\section{Conclusions}

The future of collaborative autonomous driving critically depends on overcoming current limitations in communication media, particularly regarding bandwidth efficiency, information completeness, and interoperability. Moreover, neglecting planning and control-level fusion significantly constrains the potential effectiveness of collaborative systems. Natural language communication offers a compelling solution, providing semantic richness, inherent adaptability, universal interoperability, and seamless integration with human-oriented traffic systems. By explicitly communicating intentions and reasoning, language-based systems bridge critical gaps between perception and decision-making, enhancing both safety and efficiency. We advocate for prioritizing research and development in natural language frameworks, recognizing that while complementary numeric and structured approaches may support specific use cases, natural language should serve as the foundational communication protocol, aligning multi-agent systems with the human-centric nature of transportation itself.

\clearpage

\bibliography{reference}

\end{document}